\documentclass{article}
\usepackage{amsmath}

\evensidemargin=\oddsidemargin
\def\Title#1#2#3{%
    \baselineskip=18pt
    \begin{center}
          {\large\bf{#1} \\ }
          \bigskip\bigskip
          {#2} \\
          {#3} \\
    \end{center}}
\long\def\Abstract#1{%
         \bigskip
         \parbox{0.93\textwidth}{%
                 \begin{center}
                       {\bf Abstract} \\
                 \end{center}
                 \medskip{\baselineskip=14pt #1}
                 \vss}
         \bigskip}

\makeatletter
\renewcommand{\section}%
 {\@startsection{section}{1}{0pt}%
  {-3.25ex plus -1ex minus -.2ex}{1.5ex plus .2ex}%
  {\vspace*{5mm}\raggedright\large\bf }}
\renewcommand{\subsection}%
 {\@startsection{subsection}{2}{0pt}%
  {-2.25ex plus -.5ex minus -.2ex}{-1.5ex plus -.2ex}{\bf }}
\renewcommand{\subsubsection}%
 {\@startsection{subsubsection}{3}{0pt}%
  {-1.25ex plus -.2ex minus -.1ex}{-1.2ex plus -.2ex}{\bf }}

\makeatother

\begin{document}

\Title{The Schr\"odinger equation for a spherically symmetric system,\\
 its structure and solutions}%
{R. I. Ayala O\~na\footnote{E-mail: {\tt ayyala@sfedu.ru}} and
T. P. Shestakova\footnote{E-mail: {\tt shestakova@sfedu.ru}}}%
{Department of Theoretical and Computational Physics,
Southern Federal University,\\
Sorge St. 5, Rostov-on-Don 344090, Russia}

\Abstract{The Wheeler -- DeWitt geometrodynamics, as the first attempt to develop a quantum theory of gravity, faces certain challenges, including the problem of time and the interpretation of the wave function. In this paper, we present the extended phase space approach to quantization of gravity as an alternative approach to the Wheeler -- DeWitt quantum geometrodynamics. For a spacetime with a nontrivial topology, the Wheeler -- DeWitt equation loses its sense, but we can derive the Schr\"odinger equation. Until now the Schr\"odinger equation was derived for systems with a finite number of degrees of freedom, and we need to generalize the procedure for field models. The simplest field model is a spherically symmetric one. We derive the integro-differential Schr\"odinger equation for this model, examine its structure, and find its solution.}

\section{Introduction}
Since the middle of the XXth century, several approaches to quantizing gravity have been proposed, including Wheeler -- DeWitt geometrodynamics, loop quantum gravity, and string theory. All the approaches have encountered conceptual challenges that remain subjects of intense debate in the scientific community.
For instance, the Wheeler -- DeWitt geometrodynamics faces the problem of time, when the time parameter disappears from the equations, making it difficult to describe the dynamics of the system. In string theory, extra spatial dimensions are introduced, that raises questions about their physical interpretation. In loop quantum gravity, spacetime is believed to be discretized at small scales, presenting challenges in reconciling this approach with the smooth continuum of spacetime at macroscopic scales.

An alternative to these approaches is the so-called extended phase space approach. It aims in quantizing general relativity as it is, without additional assumption about multi-dimensionality or discreteness of spacetime, but taking into account the features of gravitational theory, such as the absence of asymptotic states, which are typical for ordinary quantum theory of non-gravitational fields. In the absence of asymptotic states, one cannot use asymptotic boundary conditions in the path integral and cannot prove gauge invariance of the theory. The Wheeler -- DeWitt equation loses its sense as the main equation of the theory expressing its gauge invariance. However, the Schr\"odinger equation maintains its fundamental meaning \cite{Shest1}.

The procedure of derivation of the Schr\"odinger equation from the pass integral had been proposed by Feynman \cite{Feynman} for the simple case of a particle in an external field, and was generalized later by Cheng \cite{Cheng}, who considered the following Lagrangian,
\begin{equation}
\label{Cheng_L}
L(x,\dot x)=\frac12 g_{ij}(x)\dot x^i\dot x^j.
\end{equation}
Here $g_{ij}(x)$ is a metric of configurational space. Cheng obtained the equation
\begin{equation}
\label{Cheng_Eq}
i\hbar\frac{\partial\psi}{\partial t}
 =-\frac{\hbar^2}2\frac1{\sqrt{g}}\frac{\partial}{\partial x^i}\left(\sqrt{g}g^{ij}\frac{\partial\psi}{\partial x^j}\right)
  +\frac{\hbar^2}6R\psi.
\end{equation}
$g$ is a determinant of the metric $g_{ij}$, $R$ is a scalar curvature of the configurational space.

A feature of Eq.(\ref{Cheng_Eq}) is the quantum correction proportional to $\hbar^2$ and the scalar curvature. We shall return to it later. For system with constraints, we need further generalization, that was done for models with a finite number of degrees of freedom in \cite{SSV1,SSV2}.

In this paper, we present for the first time the derivation of the Schr\"odinger equation for a system with infinite number of degrees of freedom. We have chosen the spherically symmetric gravitational model because of its relative simplicity. Although spherically symmetric spacetime does posses asymptotic states, its topology is not trivial if we consider a black hole. Then, there exist an event horizon, and one needs two reference system to get the whole picture of spacetime. The extended phase space approach to quantization of gravity enables one to describe a situation when various reference frames are introduced in different spacetime regions \cite{Shest2}, and, in the present paper, we develop its mathematical formalism.

In this case, the interval can be written as
\begin{equation}
\label{interval}
ds^2=(N^2-N_r V)dt^2-2N_r V^2 dt dr-V^2 dr^2-W^2(d\theta^2+\sin \theta ^2 d\varphi^2).
\end{equation}
To simplify the further calculations, we have chosen $N_r=0$ from the beginning. The gravitational action is (up to the constant $\kappa$ that can be made equal to one):
\begin{equation}
\label{grav_act}
S_{(grav)}=\kappa\int dt\int\limits_0^{\infty}dr\left(\frac{\dot V\dot W W}N
 +\frac{V\dot W^2}{2N}+U\right);
\end{equation}
\begin{equation}
\label{U_pot}
U=\frac{N''W^2}V+\frac{2NWW''}V-\frac{N'V'W^2}{V^2}+\frac{2N'WW'}V-\frac{2NV'WW'}{2W^2}+\frac{N(W')^2}V-NV.
\end{equation}
From here on, the prime and the dot represent the derivative of the generalized coordinate with respect to $r$ and $t$, respectively. We shall use the effective action, as it is normally done in quantum field theory, with the gauge fixing and ghost terms:
\begin{equation}
\label{S_eff}
S_{(eff)}=S_{(grav)}+S_{(gf)}+S_{(ghost)};
\end{equation}
\begin{equation}
\label{S_gf}
S_{(gf)}=\int dt\int\limits_0^{\infty}dr\,\pi\left(\dot N-\frac{\partial f}{\partial V}\dot V
 -\frac{\partial f}{\partial W}\dot W\right);
\end{equation}
\begin{equation}
\label{S_ghost}
S_{(ghost)}=\int dt\int\limits_0^{\infty}dr\,\dot{\bar\theta}N\dot\theta.
\end{equation}

It is convenient to present the effective action in the form
\begin{equation}
\label{S_eff1}
S_{(eff)}=\int dt\int\limits_0^{\infty}dr\left[\frac12 g_{\mu\nu}\dot Q^{\mu}\dot Q^{\nu}-U
 +\pi\left(\dot N-f_{,A}\dot Q^A\right)\right],
\end{equation}
where $g_{\mu\nu}(Q^\alpha)$ is the metric of extended configurational space; $Q^\alpha=(Q^0,Q^\mu)$,
$Q^\mu=(Q^A, \theta^a)$; $Q^A=(V, W)$ are physical degrees of freedom, $\theta^a=(\bar\theta,\theta)$ are ghost variables. $Q^0=N$ is the lapse function; the gauge condition
\begin{equation}
\label{gc}
\dot N-\frac{\partial f}{\partial V}\dot V-\frac{\partial f}{\partial W}\dot W=0
\end{equation}
can be rewritten as
\begin{equation}
\label{gc1}
\dot Q^0-f_{,A}\dot Q^A=0
\end{equation}
and the comma denotes a derivative with respect to a corresponding variable. We can consider (\ref{gc1}) as a differential form of the gauge condition $Q^0=f(Q^A)+k$, $k=\rm{const}$. $g_{\mu\nu}$ is a block matrix,
\begin{equation}
\label{g_munu}
g_{\mu\nu}=\left(\begin{array}{cc}
g_{AB} & 0\\
0 & \gamma_{ab}
\end{array}\right);
\end{equation}
\begin{equation}
\label{g_AB}
g_{AB}=\left(\begin{array}{cc}
0 & \displaystyle\frac WN\\
\displaystyle\frac WN & \displaystyle\frac VN
\end{array}\right);
\quad
\gamma_{ab}=\left(\begin{array}{cc}
0 & N\\
-N & 0
\end{array}\right).
\end{equation}

The subspace with the metric $g_{AB}$ is a space of all spherically symmetric metrics determined by the field functions $N$, $V$, $W$. One can consider this space as a fiber bundle with the superspace of all possible spherically symmetric geometries of Wheeler as a base space. On the other hand, the extended configurational space with the metric $g_{\mu\nu}$ is a superspace in the sense that it includes anticommutative (Grassmannian) variables $\bar\theta$, $\theta$. One should not be confused with these two notions of superspace.

In Section 2, we describe the derivation of the Schr\"odinger equation for our field model. In Section 3, we discuss the quantum correction that is proportional to the curvature of the extended configurational space. In Section 4, we present an example of solution to the Schr\"odinger equation. Some conclusions are given in Section 5.

\section{Derivation of the Schr\"odinger equation}
We shall start from the equation
\begin{eqnarray}
\label{Feyn_Eq}
\Psi[Q(r,t+\epsilon)](t+\epsilon)
&=&\lim_{\epsilon\rightarrow 0}
 \int\exp\left(\frac i{\hbar}S[Q(t+\epsilon),Q(t)]\right)\Psi[Q(r,t)](t)\nonumber\\
&\times &\prod\limits_r\frac1{C(r)} M(Q(r,t))d\pi(r,t)\prod\limits_\alpha dQ^\alpha(r,t).
\end{eqnarray}
Eq.(\ref{Feyn_Eq}) is an analog of (18) from \cite{Feynman}. In \cite{Feynman}, $\psi(x(t),t)$ is the wave function of the mechanical system at the time moment $t$. However, in our case $\Psi[Q(r,t)](t)$ is a functional of the field functions. It is defined on extended configurational space with the coordinates
$Q^\alpha=(N,V,W,\bar\theta,\theta)$. The functional is explicitly dependent on time. $S[Q(t+\epsilon),Q(t)]$ is the approximation of the effective action at the time interval $[t+\epsilon , t]$. We also assume that the measure $M(Q(r,t))$ is not trivial.

The first step is to approximate the action $S[Q(t+\epsilon),Q(t)]$ using the equations of motion. Following to the method of \cite{Cheng}, we write the equations of motion in the form:
\begin{eqnarray}
\label{Q_Eq}
\ddot Q^A&=&-\Gamma_{BC}^A\dot Q^B\dot Q^C-\Gamma_{bc}^A\dot\theta^b\dot\theta^c+K^A;\\
\label{tet_Eq}
\ddot\theta^a&=&-\Gamma^a_{Ba}Q^B\theta^a.
\end{eqnarray}
Here, $\Gamma_{\mu\nu}^{\lambda}$ are the generalization of Christoffel symbols for superspaces \cite{AN1},
\begin{equation}
\label{Gamma}
\Gamma_{\mu\nu}^{\lambda}
 =\frac12 (-1)^{P(\nu)P(\lambda)}\left[(-1)^{P(\nu)P(\rho)}\partial_{\nu}g_{\mu\rho}
  +(-1)^{P(\mu)+P(\nu)+P(\mu)(P(\nu)+P(\rho))}\partial_{\mu}g_{\nu\rho}-\partial_{\rho}g_{\mu\nu}\right]g^{\rho\lambda},
\end{equation}
$P(\mu)$ is the parity of coordinates in the superspace: the parity is equal to 0 for commuting coordinates and 1 for anticommuting coordinates. The vector $K^A$ is determined by the potential (\ref{U_pot}). It can be demonstrated that $K^A$ does not give contribution to the form of the Schr\"odinger equation.

For generalized velocities, we use the expansions
\begin{equation}
\label{gen_vel}
\dot Q^\alpha=\frac 1{\epsilon}q^\alpha+\frac{\epsilon}2\ddot Q^\alpha
 -\frac{\epsilon^2}6\dddot Q^\alpha+\ldots;
\end{equation}
\begin{equation}
\label{q}
q^\alpha(r)=Q^\alpha(r,t+\epsilon)-Q^\alpha(r,t).
\end{equation}
$\ddot Q^\mu$, $\dddot Q^\mu$ should be expressed by means of the equations of motion.

Approximation of the gauge condition is
\begin{equation}
\label{gc_appr}
\dot Q^0-f_{,A}\dot Q^A=\frac 1{\epsilon}\left(q^0-f_{,A}q^A+\frac12 f_{,AB}q^A q^B-\ldots\right).
\end{equation}

In the integral (\ref{Feyn_Eq}), we make the replacement of variables. As a result, Eq.(\ref{Feyn_Eq}) reads:
\begin{eqnarray}
\label{Feyn_Eq1}
\Psi[Q(r,t+\epsilon)](t)+\epsilon\frac{\partial\Psi}{\partial t}
&=&\lim_{\epsilon\rightarrow 0}
 \int\exp\left(\frac i{\hbar}S[Q(t+\epsilon),Q(t)]\right)\nonumber\\
&\times &\Psi[Q(r,t+\epsilon)-q](t)
 \prod\limits_r\frac1{C(r)}M(Q(r,t+\epsilon)-q)d\pi(r)\prod\limits_\alpha dq^\alpha(r).
\end{eqnarray}

After some calculations, we get the following approximation of the action:
\begin{eqnarray}
\label{S_approx}
\frac i{\hbar}S[Q(t+\epsilon),Q(t)]
&=&\frac i{\hbar}\int\limits_0^{\infty}dr\left(\frac1{2\epsilon}\left[g_{AB}q^Aq^B
 -g_{AD}\Gamma_{BC}^D q^A q^B q^C
 +\frac14 g_{EF}\Gamma_{AB}^E\Gamma_{CD}^F q^A q^B q^C q^D\right.\right.\nonumber\\
&+&\left.\frac13 g_{AE}\left(\Gamma_{FB}^E\Gamma_{CD}^F
 +\Gamma_{BC,D}^E\right)q^A q^B q^C q^D+\ldots\right]-\epsilon U(Q)\nonumber\\
&+&\pi\left[q^0-f_{,A}q^A+\frac12 f_{,AB}q^A q^B+\ldots\right]
 +\frac1{\epsilon}Q^0\bar\theta\theta\left[1-\Gamma_{Aa}^a q^A
 +\frac12 g_{AB}\gamma^{ab}\Gamma_{ab}^B q^A\right.\nonumber\\
&+&\frac1{12}\left(5\Gamma_{Aa}^a\Gamma_{Bb}^b
 +4\Gamma_{Aa,B}^a+2\Gamma_{Ca}^a\Gamma_{AB}^C
 -3g_{CD}\gamma^{ab}\Gamma_{ab}^D\Gamma_{AB}^C
 -2g_{AC}\gamma^{ab}\Gamma_{ab}^D\Gamma_{BD}^C\right.\nonumber\\
&-&\left.\left.\left.2g_{AC}\gamma^{ab}\Gamma_{ab,B}^C
 -2g_{AC}\gamma^{ab}\Gamma_{ab}^C\Gamma_{Bc}^c\right)q^A q^B
 +\ldots\right]\vphantom{\frac12}\right).
\end{eqnarray}

Now we expand the measure:
\begin{equation}
\label{meas}
M(Q(r,t+\epsilon)-q)
=M(Q(r,t+\epsilon))-\frac{\partial^2M}{\partial Q^\alpha(r)}q^\alpha(r)
 +\frac12\frac{\partial^2M}{\partial Q^\alpha(r)\partial Q^\beta(r)}q^\alpha(r)q^\beta(r)+\ldots
\end{equation}
The measure is assumed to be dependent on $N$, $Q^A$, but not on the ghosts.

We expand $\Psi$ remembering that it is a functional:
\begin{eqnarray}
\label{Psi_exp}
\Psi[Q(r,t+\epsilon)-q]
&=&\Psi[Q(r,t+\epsilon)]
 -\int\limits_0^{\infty}dr'\frac{\delta\Psi}{\delta Q^\alpha(r')}q^\alpha(r')\nonumber\\
&+&\frac12\int\limits_0^{\infty}\!\!\int\limits_0^{\infty}\!dr'dr''
 \frac{\delta^2\Psi}{\delta Q^\alpha(r')\delta Q^\beta(r'')}q^\alpha(r')q^\beta(r'')+\ldots
\end{eqnarray}
We use the ordering rule: we write $\bar\theta$ on the left, and $\theta$ on the right. Correspondingly, we take left derivatives with respect to $\bar\theta$ and right derivatives with respect to $\theta$. This enables one to avoid needless multipliers $(-1)$ as a result of commuting Grassmannian variables.

The main distinction of our field model from the model with a finite number degrees of freedom is that the approximation of the action (\ref{S_approx}) is an integral over $r$. In (\ref{Feyn_Eq1}), we should integrate over $d\pi(r)\prod\limits_\alpha dq^\alpha(r)$ at each point $r$. To do it, we present
$\exp\left(\displaystyle\frac i{\hbar}S[Q(t+\epsilon),Q(t)]\right)$ as the product of infinite number of exponentials of the integrand in (\ref{S_approx}), each one being taken at some point $r$.
\begin{equation}
\label{prod_exp}
\exp\left(\frac i{\hbar}S[Q(t+\epsilon),Q(t)]\right)
 =\exp\left(\frac i{\hbar}\int\limits_0^{\infty}dr I(r)\right)
 =\prod\limits_r \exp\left(\frac i{\hbar}I(r)\right).
\end{equation}
Each exponential should be inserted under the sign of the product over $r$, $\prod\limits_r\{\ldots\}$ in (\ref{Feyn_Eq1}). Then, the product should be multiplied by each term of the wave functional expansion (\ref{Psi_exp}).

The second term in the right hand side of (\ref{Psi_exp}) (the integral over $r'$) can be presented as the limit of an infinite sum, each summand of the sum being dependent on $r'$.
\begin{equation}
\label{inf_sum}
\int\limits_0^{\infty}dr'\frac{\delta\Psi}{\delta Q^\alpha(r')}q^\alpha(r')
 =\lim_{\Delta r'\rightarrow 0}
  \sum\limits_{r'}\frac{\delta\Psi}{\delta Q^\alpha(r')}q^\alpha(r')\Delta r'.
\end{equation}
Each term in the sum (\ref{inf_sum}), being multiplied by the product $\prod\limits_r\{\ldots\}$, gives a contribution to (\ref{Feyn_Eq1}) when $r'=r$.

As for the third term in the right hand side of (\ref{Psi_exp}) (the double integral over $r'$ and $r''$), let us note, first of all, that it would give a contribution only when $r'=r''$ and can be reduced to a single integral over $r'$. Then, we can repeat that was said above concerning the second term.

After that, we can take integrals in (\ref{Feyn_Eq1}). The integration over $\pi$ gives a $\delta$-function of the approximated gauge condition. The integration over $q^0$ leads to a replacement of $q^0$ with $\left(f_{,A}q^A-\displaystyle\frac12 f_{,AB}q^A q^B\right)$. Now we should expand all the exponentials in series, except $\exp\left(\displaystyle\frac i{2\hbar\epsilon}\int dr g_{AB}q^A q^B\right)$. All these series should be multiplied together. The integration over ghosts should be made in accordance with the rules for Grassmannian variables. The integration over physical degrees of freedom can be made using the so-called Gaussian quadratures:
\begin{eqnarray}
\label{Gauss0}
&&\int\prod_C dq^C(r)
 \exp\left[\frac i{2\hbar\epsilon}\int\limits_0^{\infty}dr g_{AB}q^A(r)q^B(r)\right]
 =\frac{2i\hbar\epsilon\pi}{\sqrt{g}};\\
\label{Gauss1}
&&\int\prod_C dq^C(r)
 \exp\left[\frac i{2\hbar\epsilon}\int\limits_0^{\infty}dr g_{AB}q^A(r)q^B(r)\right]
  q^{D_1}q^{D_2}\ldots q^{D_{2m}}
 =\frac{2i\hbar\epsilon\pi}{\sqrt{g}}(i\hbar\epsilon)^m
 \left[g^{D_1D_2}g^{D_3D_4}\ldots g^{D_{2m-1}D_{2m}}\right.\nonumber\\
&&\hspace{3em}+\left.\mbox{terms with other possible permutations of } D_1,D_2,\ldots,D_{2m}\right].
\end{eqnarray}

After integration, we should go to the limit $\Delta r\rightarrow 0$, what gives the integral from $0$ to $\infty$ of Hamiltonian density in the right hand side of the Schr\"odinger equation.

As a result, to the zeroth order in $\epsilon$, we obtain the measure (up to a numerical constant that can be absorbed by the multiplier $C(r)$):
\begin{equation}
\label{meas_M}
M=\frac{\sqrt{g}}N=\frac W{N^2}.
\end{equation}
In the first order in $\epsilon$, we get the Schr\"odinger equation:
\begin{equation}
\label{math_Sch}
i\hbar\frac{\partial\Psi}{\partial t}
 =\int\limits_0^{\infty}dr\left[-\frac{\hbar^2}{2M}\frac{\delta}{\delta Q^\alpha}
  \left(MG^{\alpha\beta}\frac{\delta}{\delta Q^\beta}\right)
 +U+\frac{\hbar^2}6 R\right]\Psi,
\end{equation}
where
\begin{equation}
\label{G_al_be}
G^{\alpha\beta}=\left(\begin{array}{ccc}
f_{,C}f^{,C} & f^{,B} & 0\\
f^{,A} & g^{AB} & 0\\
0 & 0 & \gamma^{ab}
\end{array}\right);
\end{equation}
$R$ is a scalar curvature of the extended configurational space.

Let us emphasised that the Schr\"odinger equation (\ref{math_Sch}) is {\it a direct mathematical consequence} of the path integral with the effective action (\ref{S_eff1}), which is considered without asymptotic boundary conditions generally accepted in quantum theory of non-gravitational fields. We shall refer to it as {\it the mathematical Schr\"odinger equation}.

\section{The quantum correction}
As in the work of Cheng \cite{Cheng}, we have obtained the quantum correction $\displaystyle\frac{\hbar^2}6 R$ (compare (\ref{Cheng_Eq}) and (\ref{math_Sch})). In \cite{Cheng}, the correction is proportional to the curvature of configurational space. As was said above, in our case, the configurational space is superspace since it includes ghosts (Grassmannian) variables $\bar\theta$, $\theta$.

In \cite{Ber1}, F. A. Berezin wrote:
\begin{quote}
``A striking coincidence of basic formulas of operator calculus in Fermi and Bose variants of the second quantization was discovered in 1961.''
\end{quote}
(See also \cite{Ber2}). Therefore, we can expect that, when one includes Grassmannian variables into the extended configurational space, the quantum correction is also proportional to its curvature, and the curvature can be calculated in accordance with uniform rules.

The emergence of the concept of supersymmetry has caused the need for generalization of Riemannian geometry to the case of superspaces. Among the first works on geometry of superspaces were those by Arnowitt and Nath \cite{AN1,AN2} and by Akulov, Volkov and Soroka \cite{AVS,AV}. Later, DeWitt wrote his book ``Supermanifolds'' \cite{DeWitt}. However, there are ambiguities in definitions of some quantities. For example, in \cite{AVS}, two ways are given to define an invariant contraction over tensor indices. The formulae for the curvature tensor and scalar in \cite{AN1} and \cite{DeWitt} do not coincide. It may be explained by that different definitions of covariant derivatives were used by different authors (they used covariant derivatives of covariant or contravariant vectors and various orders of multipliers; it all matters in the case of superspace). The scalar curvature we obtained coincides with that in \cite{AN1}. It can be calculated according to the following equations:
\begin{equation}
\label{curv_Riem}
R_{\mu\nu\lambda}^\rho
=-\Gamma_{\mu\lambda,\nu}^\rho+(-1)^{P(\nu)P(\lambda)}\Gamma_{\mu\nu,\lambda}^\rho
-(-1)^{P(\lambda)(P(\rho)+P(\sigma))}\Gamma_{\mu\lambda}^\sigma\Gamma_{\sigma\nu}^\rho
+(-1)^{P(\nu)(P(\lambda)+P(\rho)+P(\sigma))}\Gamma_{\mu\nu}^\sigma\Gamma_{\sigma\lambda}^\rho;
\end{equation}
\begin{equation}
\label{Ricci}
R_{\mu\nu}=(-1)^{P(\lambda)}R_{\mu\nu\lambda}^\lambda;
\end{equation}
\begin{equation}
\label{curv_scal}
R=(-1)^{P(\nu)}g^{\nu\mu}R_{\mu\nu},
\end{equation}
and the Christoffel symbols are given by (\ref{Gamma}). Our result witnesses for the approach to geometry of superspaces developed by Arnowitt and his collaborators. We believe that, taking into account the availability of different approaches, our result can be a criterion of correctness of how the superspace geometry should be constructed.

\section{Example of a solution to the Schr\"odinger equation}
The general solution to the mathematical Schr\"odinger equation (\ref{math_Sch}) is
\begin{equation}
\label{gen_sol}
\Psi[Q^\alpha](t)=\int dk\,\Phi[Q^A](t)
 \delta\left(N-f(Q^A)-k\right)\left(\bar\theta(t)+i\theta(t)\right).
\end{equation}
Here, $\Phi[Q^A](t)$ is a functional that is dependent only on physical degrees of freedom, $Q^A$. The substitution of (\ref{gen_sol}) to (\ref{math_Sch}) gives the equation for $\Phi[Q^A](t)$:
\begin{equation}
\label{phys_Sch}
i\hbar\frac{\partial\Phi}{\partial t}
 =\int\limits_0^{\infty}dr\left[-\frac{\hbar^2}{2M}\frac{\delta}{\delta Q^A}
  \left(Mg^{AB}\frac{\delta}{\delta Q^B}\right)
 +\left. U+\frac{\hbar^2}6 R\right]\Phi\right|_{N=f(Q^A)+k}.
\end{equation}
This equation is of the main interest for us, since it describes physical degrees of freedom. Thus, we shall refer to it as {\it the physical Schr\"odinger equation}. As in Eq.(\ref{math_Sch}), we have obtained the equation in functional derivatives which are included in the integrand in the right hand side of (\ref{phys_Sch}). In the explicit form it looks like
\begin{eqnarray}
\label{phys_Sch_expl}
i\hbar\frac{\partial\Phi}{\partial t}&=&\int\limits_0^{\infty}dr\left\{\hbar^2
 \left[\frac{Vf}{2W^2}\frac{\delta^2\Phi}{\delta V^2}
 -\frac fW\frac{\delta^2\Phi}{\delta V\delta W}\right.\right.
 +\left(\frac 1{2W}\frac{\partial f}{\partial W}-\frac V{2W^2}\frac{\partial f}{\partial V}
 +\frac f{2W^2}\right)\frac{\delta\Phi}{\delta V}\nonumber\\
&+&\left.\frac 1{2W}\frac{\partial f}{\partial V}\frac{\delta\Phi}{\delta W}\right]+U\Phi
 +\hbar^2\left[\frac{7V}{12fW^2}\left(\frac{\partial f}{\partial V}\right)^2
 -\frac V{2W^2}\frac{\partial^2f}{\partial V^2}\right.\nonumber\\
& +&\frac 1W\frac{\partial^2f}{\partial V\partial W}
 -\frac 7{6fW}\frac{\partial f}{\partial V}\frac{\partial f}{\partial W}
 -\left.\left.\frac 1{2W^2}\frac{\partial f}{\partial V}\right]\Phi\right\}.
\end{eqnarray}

At the present level of investigations, we do not have well developed methods of solving this quite complicated equation. Now we attempt to find a solution to the stationary Schr\"odinger equation corresponding to (\ref{phys_Sch_expl}) in some limited class of functions, ignoring the quantum correction proportional to $\hbar^2$. We consider the so-called ``Schwarzschild gauge'' $N=\displaystyle\frac 1V$, that leads to the Schwarzschild solution to the classical Einstein equations, and put $W(r)=r$ (it means that the radial coordinate $r$ is chosen so that the circumference length is equal to $2\pi r$). Then, the functional $\Phi$ will depend only on $V(r)$. In this case, the stationary Schr\"odinger equation reads:
\begin{equation}
\label{phys_Sch_exp2}
E\Phi[V]=\int\limits_0^{\infty}dr\left[\frac{\hbar^2}{2r^2}\left(\frac{\delta^2\Phi}{\delta V^2}
 +\frac 2V\frac{\delta\Phi}{\delta V}\right)+U\Phi\right].
\end{equation}
The potential looks like
\begin{equation}
\label{U_pot_1}
U=\frac{3(V')^2r^2}{V^4}-\frac{V''r^2}{V^3}-\frac{4V'r}{V^3}+\frac1{V^2}-1.
\end{equation}

To ensure convergence of the integral, we require that $U=0$. The requirement $U=0$ gives a differential equation defining a class of admissible functions $V(r)$. One can check that
\begin{equation}
\label{V_class}
V(r)=\frac1{\sqrt{1+\frac Ar+\frac B{r^2}}}.
\end{equation}
It is remarkable that (\ref{V_class}) gives the Schwarzschild solution when $A=-2M$ and $B=0$ and the Reissner -- Nordstr\"om solution when $A=-2M$ and $B=Q_{bh}^2$ \cite{HE} ($M$ is the mass of a central body, $Q_{bh}$ is its charge).

The functional $\Phi[V]$ is sought in the form
\begin{equation}
\label{Phi_1}
\Phi[V]=\int\limits_0^{\infty}dr e^{-r}\phi(V(r)).
\end{equation}
The substitution (\ref{Phi_1}) into (\ref{phys_Sch_exp2}) gives
\begin{equation}
\label{phys_Sch_exp3}
E\int\limits_0^{\infty}dr e^{-r}\phi(V(r))
 =\int\limits_0^{\infty}dr e^{-r}\frac{\hbar^2}{2r^2}\left(\frac{\partial^2\phi}{\partial V^2}
 +\frac 2V\frac{\partial\phi}{\partial V}\right),
\end{equation}
and we obtain the differential equation at each point $r$:
\begin{equation}
\label{phi_Eq}
\frac{\partial^2\phi}{\partial V^2}+\frac 2V\frac{\partial\phi}{\partial V}-\frac{2Er^2}{\hbar^2}\phi(V)=0.
\end{equation}
Considering $\lambda(r)=\sqrt{2E}\displaystyle\frac r{\hbar}$ as a quantity not depending on $V$, we can write the solution to Eq.(\ref{phi_Eq}):
\begin{equation}
\label{gen_sol_phi}
\phi(V)=\frac 1V\left(C_1e^{\lambda V}+C_2e^{-\lambda V}\right).
\end{equation}
Choosing a not-growing solution, we finally have:
\begin{equation}
\label{phys_sol1}
\Phi[V]
 =\int\limits_0^{\infty}dr\frac C{V(r)}\exp\left[-r\left(1+\displaystyle\frac{\sqrt{2E}}{\hbar}V(r)\right)\right].
\end{equation}

In fact, taking into account that $V(r)$ belongs to the class of functions (\ref{V_class}), the functional $\Phi[V]$ becomes a function of two parameters $A$ and $B$ which determine the mass and charge of a central body.

\section{Conclusions}
In the present paper, we consider the procedure of derivation of the Schr\"odinger equation for spherically symmetric gravitational model which is the simplest example of a field model. This procedure generalizes methods that were used earlier for models with finite number of degrees of freedom. We believe that the mathematical technique we developed here can be applicable to any field model.

As early in the work of Cheng \cite{Cheng}, we obtain the quantum correction proportional to the scalar curvature of configurational space. Since we used the effective action and considered the path integral without asymptotic boundary conditions, the configurational space includes ghost (Grassmannian) variables; in this sense, it is superspace. There are some ambiguities in formulae defining important objects of superspace geometry, namely, the ambiguities in grading factors $(-1)^{P(\ldots)}$. To avoid these ambiguities, one has to choose correct definition of such objects as Christoffel symbols, curvature tensors, etc. We hope that our experience may be useful for supersymmetric extensions of general relativity and in cases when the action includes not only ghost fields, but also anticommuting fermionic fields.

In field models, a solution to the Schr\"odinger equation is not a function but a functional of field variables. In the right hand side of the equation, the Hamilton operator is an integral of Hamiltonian density which includes functional derivatives. Thus, finding a solution is a difficult task. We have found a solution based on the assumption that the potential (\ref{U_pot}) must be equal to zero on admissible function $V(r)$. The sought-for functional turned out to be a function of two parameters, characterizing the mass and charge of a central body (or a black hole).

\end{document}